\documentclass{aa}
\usepackage{amssymb}
\usepackage{graphicx}
\usepackage{hyperref}

\def\asec{\ifmmode ^{\prime\prime}\else$^{\prime\prime}$\fi}

\def\degs{\ifmmode ^{\circ}\else$^{\circ}$\fi}
\def\amin{\ifmmode ^{\prime}\else$^{\prime}$\fi}
\def\asec{\ifmmode ^{\prime\prime}\else$^{\prime\prime}$\fi}
\def\fm{\hbox{$.\!\!^{\rm m}$}}            
\def\fdg{\hbox{$.\!\!^\circ$}}          
\def\farcs{\hbox{$.\!\!^{\prime\prime}$}}  

\def\psr{PSR~B0833$-$45}
\def\js{J$_{\rm s}$}
\def\degs{\ifmmode ^{\circ}\else$^{\circ}$\fi}
\def\amin{\ifmmode ^{\prime}\else$^{\prime}$\fi}
\def\farcm{\hbox{$.\mkern-4mu^\prime$}}

\def\hst{{\sl HST\/}}
\def\rxte{{\sl RXTE\/}}

\def\chandra{{\sl Chandra\/}}

\def\widerul{\vrule height 2.5ex width 0ex depth 0ex}
\def\wideru{ \vrule height 2.8ex   width 0ex depth 0ex}
\def\widerul{\vrule height 2.5ex width 0ex depth 0ex}

\def\twolines#1#2{
\renewcommand{\arraystretch}{0.8}
\begin{tabular}{@{}c@{}}
#1 \vrule height3.2ex width0ex \\ #2 \\[.7ex]
\end{tabular}
}

\def\fourlines#1#2#3#4{
\renewcommand{\arraystretch}{0.8}
\begin{tabular}{@{}c@{}}
#1 \vrule height3.2ex width0ex \\ #2 \\[-.0ex]
\hline\\[-3.4ex]
#3 \vrule height3.2ex width0ex \\ #4 \\[.7ex]
\end{tabular}
}

\begin{document}

\title{The Vela Pulsar in the Near-Infrared\thanks{Based 
on observations collected at the European Southern Observatory, Paranal, Chile
(ESO Programme 66.D-0568).}}

\author{%
Yu.A.~Shibanov\inst{1}
\and 
A.B.~Koptsevich\inst{1}
\and 
J.~Sollerman\inst{2} 
\and
P.~Lundqvist\inst{2}
}

\institute{%
Ioffe Physical Technical Institute, Politekhnicheskaya 26, St.\ Petersburg, 194021, Russia 
\and 
Stockholm Observatory, AlbaNova, Department of Astronomy, SE-106 91 Stockholm, Sweden 
} 

\date{Received  30 January 2003 / Accepted 25 April 2003 } 
\authorrunning{Shibanov et al.}
\titlerunning{The Vela Pulsar in the Near-Infrared } 
\offprints{Yuri~Shibanov, \newline {\tt shib@astro.ioffe.rssi.ru}}

\abstract{We report on the first detection of 
the  Vela pulsar in the near-infrared  
 with the VLT/ISAAC in the \js\ and H bands. The pulsar magnitudes are 
$J_{\rm s}=22.71\pm0.10$ and $H=22.04\pm0.16$.  
We compare our results with the available  
multiwavelength data and show that the dereddened  
phase-averaged optical spectrum  of the pulsar can be fitted with a power law 
$F_\nu \propto \nu^{-\alpha_\nu}$  with  $\alpha_\nu = 0.12\pm0.05$,    
assuming the color excess $E_{B-V}=0.055\pm0.005$  
based on recent  spectral 
fits of the emission of the Vela pulsar and its supernova remnant in X-rays. 
The negative slope of the pulsar spectrum is 
different from the positive slope observed 
over a wide optical range in the young Crab pulsar spectrum. 
The near-infrared  part of the Vela spectrum appears to have
the same slope as  the  phase-averaged spectrum in the high energy 
X-ray tail, obtained in the $2-10$~keV range with the \rxte. 
Both of these spectra 
can be fitted with a single power law suggesting their common origin. 
Because the phase-averaged \rxte\ spectrum  
in this range is dominated by the second X-ray peak 
of the pulsar light curve, 
coinciding with the second main peak of its optical pulse 
profile, we suggest that this optical peak can be redder than the first one.   
We also detect two faint extended  structures  
in  the 1\farcs$5-3$\farcs1 vicinity of the pulsar, projected on and 
aligned with the south-east jet and the inner arc of the pulsar 
wind nebula, detected  in X-rays with \chandra. 
We discuss their possible association with the nebula.             

\keywords{Infrared:  general --  pulsars: individual:  Vela pulsar --
stars: neutron} }

\maketitle

\section{Introduction}

 The Vela pulsar was firmly identified 
 in the optical range by the 
 detection of optical pulsations with the radio pulsar period   
(\cite{wal77}; \cite{man78}).
 The identification has been 
further confirmed
by measuring the proper motion and 
 parallax of the radio pulsar and its 
optical counterpart  
(\cite{leg00}; \cite{Caraveo}),
by broad-band photometry 
revealing peculiar colors of the counterpart 
typical for the optical emission   
  of rotation powered pulsars    
(\cite{las76}; \cite{mign} and refs.\ therein),    
and by the high polarization of the optical 
emission (\cite{wagner}).
The Vela pulsar is an intermediate age, 
$\sim 10^4$~yr, isolated  neutron star (NS). 
Its parameters are listed in Table~\ref{t:param}.  

In comparison with older pulsars detected 
in the optical range (see, e.g., 
\cite{Mignani2}),
the Vela pulsar  with  V= 23\fm6 is brighter 
by at least $1-2$ stellar magnitudes.   
However, available spectral information 
 on its optical emission  
 has been limited to broad-band 
 UBVRI photometry (\cite{mign}), which suggests            
a  flat optical spectrum  typical for young Crab-like pulsars.  
 This is in contrast to the  middle-aged, $\sim 10^5$~yr,  
pulsars PSR B0656$+$14 and Geminga  whose broad-band spectra are 
 less monotonous, and to the strong excess in the near infrared (IR) 
part of the spectrum of PSR B0656$+$14 (\cite{kopts01}). 
This may be an evidence of spectral evolution of the optical emission 
with pulsar age.  The young Crab pulsar shows no excess in the IR. 
Thus, the extension of the spectrum of the Vela pulsar towards the IR 
 is useful to determine whether the optical properties of this pulsar 
are closer to those of younger or middle-aged 
NSs, and to get additional constraints on the pulsar 
spectral evolution with age.    

The Vela pulsar has also been studied in the high energy range, 
from soft X-rays to $\gamma$-rays (e.g., \cite{pzs01}; \cite{hsg02} and refs.\ therein). 
Its multiwavelength spectrum  is presumably nonthermal with different 
slopes in different high energy ranges.  An exception is the soft X-ray range 
where a strong excess over a power law\footnote{Hereafter PL, 
$F_\nu\propto \nu^{-\alpha_\nu}$.} background is believed to be due to 
the thermal emission from  the surface of the NS (\cite{og93}; \cite{pzs01}). 
Different slopes imply different nonthermal radiation mechanisms at work 
in the magnetosphere of the pulsar (e.g., synchrotron, curvature,  
inverse Compton scattering radiation,  etc.). These mechanisms are involved 
in different ways in the two competing models for the non-thermal emission 
of pulsars, the ``polar cap''  model (e.g., \cite{dh96}) and the ``outer gap'' 
model  (e.g., \cite{chr86}; \cite{rom96}).
\begin{table*}[t]
\caption{Parameters of the Vela pulsar (\psr; \cite{Taylor}, unless specified otherwise).}
\label{t:param}
\begin{tabular}{ccccccccccc}
\hline\hline
\multicolumn{6}{c}{Observed}&&\multicolumn{4}{c}{Derived} \widerul\\
\cline{1-6}\cline{8-11}
$P$ &
$\dot P$ &  
$\mu_\alpha\cos(\delta)$,\ \  $\mu_\delta^a$ &
$\pi^a$ &
$l$, $b^b$ &
$D\!M^c$ &&
$\tau^d$ &
$B$ &
$\dot E$ & 
$d^{a}$ \wideru \\
ms &
$10^{-15}$ &
mas yr$^{-1}$ &
mas &&
cm$^{-3}$ pc &&
Myr & 
G  &
erg s$^{-1}$ &
pc \widerul \\ 
\hline 
89 &
$125$ &
\twolines{$-37.2\pm1.2$}{$28.2\pm1.3$} &
$3.4\pm0.7$ &
\twolines{263\fdg6}{$-$2\fdg8} &
$68.2$ &&
0.011 &
$3.38 \times 10^{12}$ &
$6.9 \times 10^{36}$ &
$294^{+76}_{-50}$ \widerul\\
\hline
\end{tabular}
\begin{tabular}{ll}
$^a$~Proper motion, parallax, and parallax-based distance (\cite{Caraveo}) & 
$^c$~Dispersion measure \wideru\\
$^b$~Galactic coordinates    &   
$^d$~Dynamical age \wideru \\ 
\end{tabular}
\end{table*}

It is not yet clear which of these competing models  
best represents 
the observations, and which radiation mechanisms 
are actually 
responsible for the observed emission in each 
band.
In this respect, observations in the IR are important to get additional 
constraints on these mechanisms and models. 
For instance, in the polar cap model IR photons, 
as well as optical ones,  can be produced by inverse Compton 
scattering of softer photons by primary and/or pair cascade 
relativistic particles in the magnetosphere.

A symmetrical Crab-like pulsar wind nebula (PWN) with a torus and 
jet structure has recently been detected in X-rays 
with the \chandra\ observatory around the Vela pulsar 
(\cite{hgh01}; \cite{pzs01}; \cite{pks01}). 
Being fainter than 
the Crab PWN, the Vela nebula has not yet been detected in 
the optical range, perhaps because the pulsar field is crowded 
by background stars. Some of the structures of the Crab PWN, identified 
in the optical, appear brighter in the near-IR range 
(\cite{sf02}),  
showing a PL spectrum with a negative slope close to that 
observed in X-rays.  There are indications that the X-ray Vela PWN 
has a PL spectrum of a similar slope 
(\cite{go02}). 
In this context, deep imaging of the Vela pulsar field in the near-IR 
might be more promising than imaging in the visual range   
for detection of the Vela PWN and for studying the  
mechanisms of the interaction between
the relativistic pulsar wind and the ambient matter. 
       
Here we report on the first detection of the Vela pulsar in the near-IR
\js\ and H bands, obtained with the VLT.  
The observations, data reduction, astrometry,  
and photometry are described in Sect.~\ref{s:obsanal}. 
The results are discussed  
 in Sect.~\ref{s:disc}
in conjunction with the data available in other spectral bands, 
and summarized in Sect.~\ref{s:sum}.

\section{Observations and data analysis}

\label{s:obsanal}

\subsection{ISAAC observations and data reduction}

The field of the Vela pulsar was observed during three nights, 
December 14 and 15, 2000, and January 5, 2001 with the 
Infrared Spectrometer And Array 
Camera\footnote{See
\href{http://www.eso.org/instruments/isaac/}
{http://www.eso.org/instruments/isaac/} 
for details on the 
instrument, filters and observational technique.} (ISAAC)
attached to the Antu Telescope (UT1) of the 
European Southern Observatory's Very Large Telescope (ESO VLT).
A log of the observations is given in Table~\ref{t:obs}.
In the SW (Short Wavelength) imaging mode, the
Rockwell Hawaii  HgCdTe 1024$\times$1024 array detector was used. 
The pixel size on the sky was 0\farcs147 and the field-of-view was 
2\farcm5$\times$2\farcm5.
The observations were performed in the 
\js\ and H bands in jitter mode, with a
jitter box size of 20\asec. 
 The \js\ filter is being established as the new standard broadband 
filter at $\approx 1.24\mu$m by most major observatories (Keck, Gemini, 
Subaru, ESO), and is photometrically more accurate than the classical 
J because it is not cut off 
by atmospheric absorption (\cite{simons}; \cite{labbe}).         
The detector integration times (DITs) were 45~s and  
13~s in the \js\ and H bands, respectively. Each observation was built up by
a number of DITs per exposure (NDITs), where we used 4 NDITs for the \js\ 
band and 6 NDITs for the H band.
The number of exposures (NEXP) in each observational block 
(ObsID) was 13 in \js\ and varied 
from 12 to 26 in the H band.
Total exposure times (NDIT$\times$DIT$\times$NEXP summed over all ObsIDs) 
were thus 7020~s and 8268~s in the \js\ and H bands, respectively.

\begin{table}[b]
\caption{Log of VLT/ISAAC observations of the Vela pulsar.}
\label{t:obs} 
\begin{tabular}{llllll}
\hline
\hline
Band &Date       &Time$^a$& Exposure &Airmass & Seeing$^b$  \\ 
     & UT        &    UT    &   s      &            & arcsec \\
\hline
\js\ &14.12.00 & 7:45     &   2340   & 1.072 & 0.7  \\
\cline{2-6}
     &15.12.00 & 6:33     &   2340   & 1.084 & 0.6 \\
     &           & 7:18     &   2340   & 1.069 & 0.5 \\
\hline
H    &15.12.00 & 8:03     &   1014   & 1.076 & 0.5 \\
     &           & 8:29     &   1092   & 1.094 & 0.5 \\
     &           & 8:54     &    936   & 1.118 & 0.7 \\
\cline{2-6}
     &05.01.01 & 6:20     &   2028   & 1.073 & 0.5 \\
     &           & 7:05     &   2028   & 1.103 & 0.4 \\
     &           & 7:51     &   1170   & 1.150 & 0.4 \\
\hline
\end{tabular} \\
\begin{tabular}{lll}
$^a$ \ Refers to the first image of the ObsID. && \\
$^b$ \  Full width at half maximum of the stellar profile.&& \\
\end{tabular}
\end{table}

\begin{figure*}
\setlength{\unitlength}{1mm}
\begin{picture}(178,160)(0,0)
\put (  0, 80)   {\includegraphics[width=85mm,bb=150 273 420 513,clip]{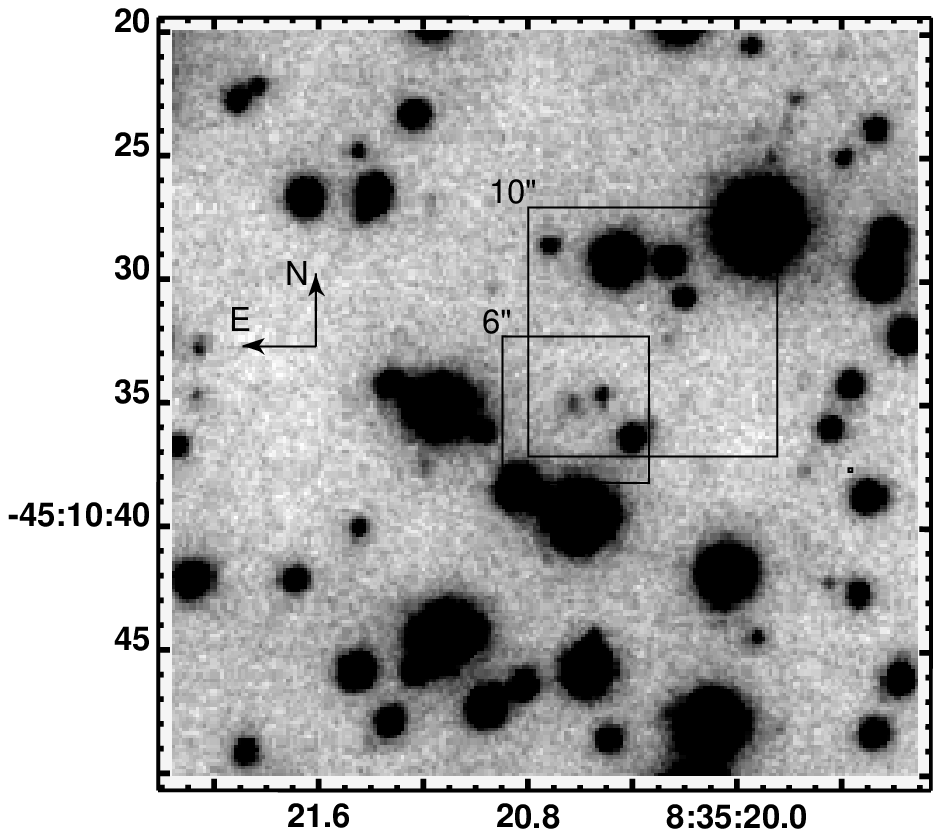}}
\put ( 90,119)   {\includegraphics[width=35.5mm,bb=220 315 392 488,clip]{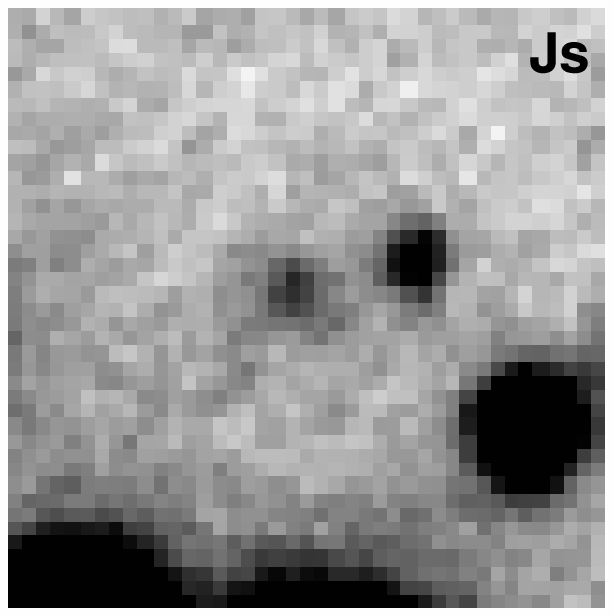}}
\put (128,117)   {\includegraphics[width=38mm,bb=55 60 522 522,clip]{h4279f1-r2.eps}}
\put ( 90, 80)   {\includegraphics[width=35.5mm,bb=220 315 392 488,clip]{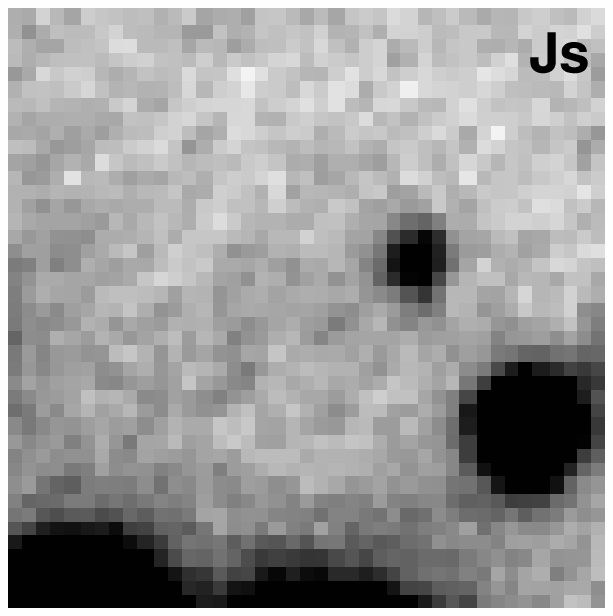}}
\put (128, 78)   {\includegraphics[width=38mm,bb=55 60 522 522,clip]{h4279f1-r4.eps}}
\put ( 90, 40)   {\includegraphics[width=35.5mm,bb=220 315 392 488,clip]{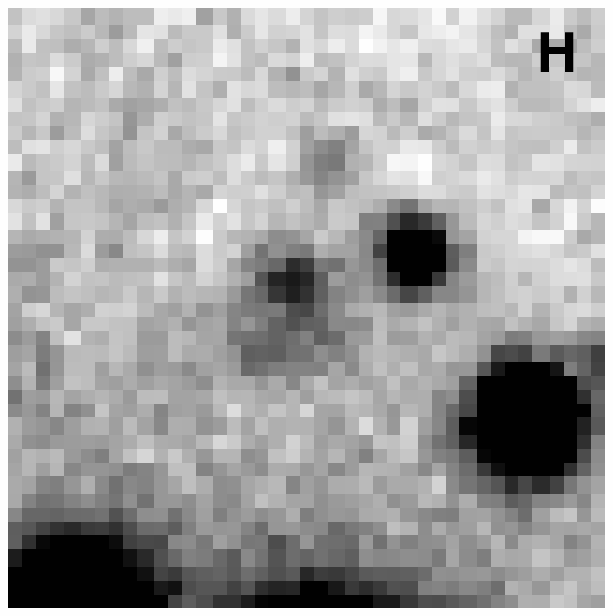}}
\put (128, 38)   {\includegraphics[width=38mm,bb=55 60 522 522,clip]{h4279f1-r6.eps}}
\put ( 90,  0)   {\includegraphics[width=35.5mm,bb=220 315 392 488,clip]{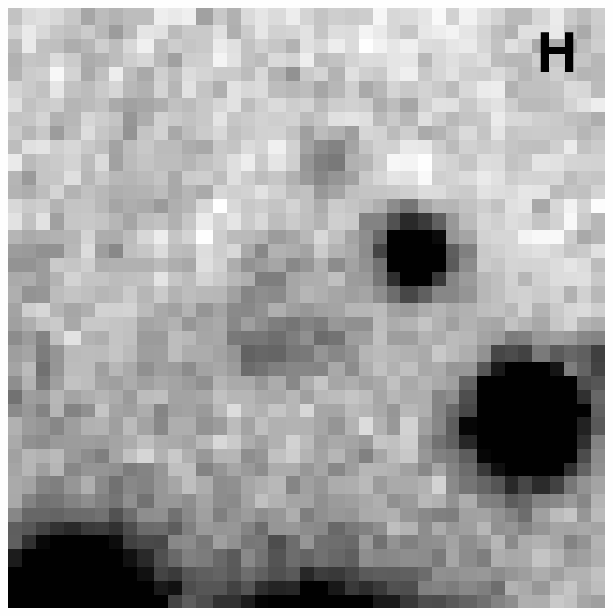}}
\put (128, -2)   {\includegraphics[width=38mm,bb=55 60 522 522,clip]{h4279f1-r8.eps}}
\put ( 0, -1)   {\parbox[b]{85mm}{\caption{%
{\bfseries\itshape Large panel:}
$30\arcsec\times30\arcsec$ overview of the
Vela pulsar field extracted from the $2\farcm5\times2\farcm5$ 
frame of the VLT/ISAAC
image in the \js\ band.
The region bounded by the 6\arcsec-box is enlarged in the {\it small
panels.\/} The same region in different bands is shown 
in Figs.~\ref{f:fors} and \ref{f:wfpc}.
The region bounded by the 10\arcsec-box is shown in Fig.~\ref{f:arc}.
{\bfseries\itshape Small panels} show the $6\arcsec\times6\arcsec$
field of view centered at the pulsar position
in the \js\ and H bands.  
Images with the pulsar counterpart subtracted
are  also shown in the lower panels for each band, and the corresponding 
contour plots are given. Reference frame for the contour plots  
shows the image scale only. 
The pulsar and nearby objects o1, o2 and o3 are marked in the H 
band contour plot. The crosses in the \js\ contour plots show the pulsar 
position as derived from the
\hst/WFPC2/F555W image 
(\cite{Caraveo})
with an uncertainty negligible for this
image scale.
The arrow in the \js\ contour plot shows the pulsar proper motion direction,  
see Sect~\ref{s:astrmorph} for details.    
\label{f:ima}
}}}
\end{picture}
\end{figure*}

The data were reduced with
the {\tt Eclipse}\footnote{See
\href{http://www.eso.org/projects/aot/eclipse/}
{http://www.eso.org/projects/aot/eclipse/}.}
and NOAO {\tt IRAF} software. 
Dark and flatfield images were prepared
using standard {\tt Eclipse} recipes.
Then each ObsID was considered separately.
Image offsets were determined using the {\tt geomap} routine and 
four field stars.
The sky background level was determined and the images were summed 
using the routine {\tt jitter} with the
parameters {\tt RejectHalfWidth = 7, RejectMin = 2, RejectMax = 4},
which were chosen based on the resulting image statistics.
The parameters of the fully reduced images for each ObsID are summarized in
Table~\ref{t:obs}. 
Finally, offsets between these images
were determined and the images were combined. Parts of the resulting
images\footnote{Images are available in
{\tt FITS} format at \href{http://www.ioffe.ru/astro/NSG/obs/vela-ir/}
{http://www.ioffe.ru/astro/NSG/obs/vela-ir/}.} containing the pulsar 
are shown in  Fig.~\ref{f:ima}.
Contour maps of these images are also presented in this figure.
Isophotes of the contour maps correspond to the levels
(in  counts) above the  background $l_n=S+n\sigma$,  where $S$  is the
mean sky value near the pulsar, $\sigma$ is the sky standard deviation
per pixel, and  $n=1,2,\ldots,6$.

\subsection{Astrometry and morphology of the pulsar field}

\label{s:astrmorph}

For astrometrical referencing of the VLT images  
we used the \hst/WFPC2 image  
obtained on January 15, 2000   
(\cite{Caraveo}).
The pulsar is clearly detected in this image. 
Positions of 11 reference stars from the image were used 
to construct the coordinate transformation between 
the \hst\ and VLT images with the IRAF routines      
{\sf geomap/geoxytran}.  
The {\sl rms\/} errors of the transformation were less than 
one third of  
the ISAAC pixel size in both RA and Dec.
The pulsar position in the ISAAC images 
at the epoch of the VLT observations
was calculated using the pulsar pixel coordinates 
in the \hst\ image and the pulsar proper motion  
(\cite{Caraveo}).
This position is marked 
by a cross in the contour plots of the \js\ images in Fig.~\ref{f:ima}.
The pulsar counterpart is clearly detected
with the offsets $-0\farcs01(5)$\footnote{Hereafter the numbers in parentheses 
are uncertainties referring to the last significant digit quoted, 
for example, 
$0.01(5)=0.01\pm0.05$, $22.04(18)=22.04\pm0.18$.}
 and 0\farcs01(2)
in RA and Dec, respectively, from this position. The errors account for the centering uncertainties 
in the ISAAC images, coordinate transformation and the pulsar proper motion uncertainties.

\begin{figure}[t]
\includegraphics[width=50mm,bb=232 323 378 470,clip]{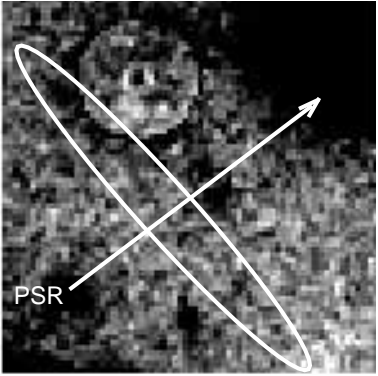}
\caption{$10\arcsec\times10\arcsec$ vicinity of the Vela 
pulsar in the \js\ band. 
Objects o2, o3, and four stars in the upper left corner of the image
are subtracted (cf.\ Fig.~1). 
The arrow shows the pulsar proper motion direction.
A faint thin elongated object oriented approximately perpendicular 
to the proper motion is seen  within the region bounded by     
the ellipse.  It coincides with the central part of 
the inner arc of the PWN 
detected in X-rays (see Sect.~\ref{s:nearby} for details).}
\label{f:arc}
\end{figure}
 
Three point-like objects, o1, o2, and o3, 
are detected in the 6\arcsec$\times$6\arcsec\ vicinity 
of the pulsar. 
They are marked in the H-band contour plot in Fig.~\ref{f:ima}, and 
their offsets from  the pulsar position 
are listed in Table~\ref{t:obj}. 
The faintest object o1 is detected only in the H band.

An extended structure is seen 
in the 1\farcs5 vicinity 
south  of the pulsar.
The structure is more pronounced in the H band, but can also be 
marginally detected
in the \js\ image, although 
with a different shape. 
The examination of the structure 
in each ObsID shows that 
its shape varies from one ObsID to another.     
To discriminate the pulsar from the extended structure,
the {\tt IRAF/DAOPHOT} software was used to 
construct a point spread function (PSF) in both bands using 10 field stars.
The pulsar spatial profile was fitted with this PSF and  
subtracted from the images. 
The subtracted images are presented in Fig.~\ref{f:ima}. 
\label{s:anal}

The  
extension of the structure in the 
\js\ band 
is aligned with  
the south-east 
counter-jet from the pulsar detected in 
X-rays with the \chandra\ observatory  
(\cite{hgh01}; \cite{pks01}). 
The X-ray counter-jet is directed opposite to the pulsar proper motion 
marked by an arrow in the \js\ contour plot in Fig.~\ref{f:ima}. 
It extends up to $\sim100$\arcsec\ from  
the pulsar. The ISAAC structure  can be a near-IR 
signature of the X-ray jet in the 2\arcsec\ vicinity of the pulsar. 
However, visual 
inspection of the 2\arcsec\ vicinity of the pulsar 
in the \chandra/HRC image does not reveal any such structure. 
It may be hidden in the complicated pulsar PSF 
profile of the HRC image.

We found  in the ISAAC images also a hint of a faint  thin 
elongated  
structure, overlapping with the central part of the
inner arc of the Vela PWN in the \chandra/ACIS images (\cite{pks01}).
The structure  is seen 
within the ellipse in Fig.~\ref{f:arc}.  
 It is aligned approximately perpendicular to the pulsar proper 
motion direction, as is the X-ray arc,  and its offset    
from the pulsar is 3\farcs1 along this direction.
The structure is  detected at only  $\sim 2\sigma$ level  
and only  in the \js\ band  (see Sect.~\ref{s:phot} for details).    
However, inspection  of each separate ObsID image (see Table~\ref{t:obs}) 
shows that the structure is absent only 
in the first \js\ image, 
which has the worst seeing, but it is present in the two other images.

\begin{figure}[t]
\setlength{\unitlength}{1mm}
\begin{picture}(80,80)(0,0)
\put (  0,38)   {\includegraphics[width=36mm,bb=175 275 435 535,clip]{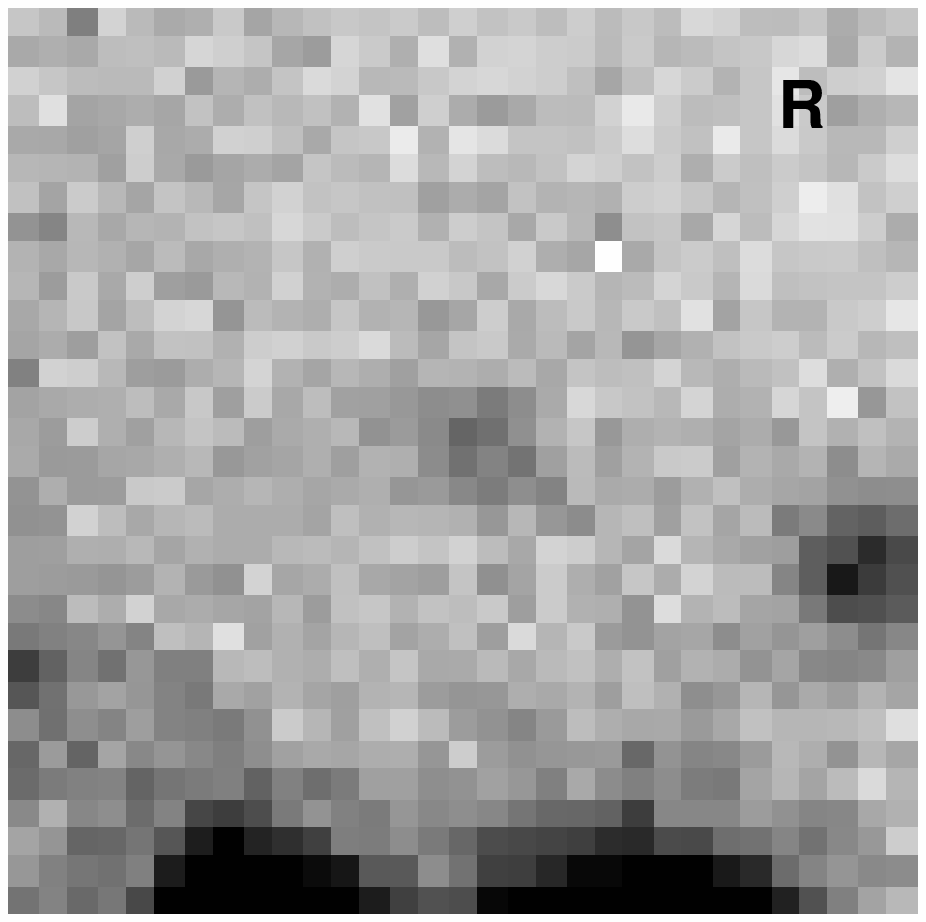}}
\put (  0, 0)   {\includegraphics[width=36mm,bb=175 275 435 535,clip]{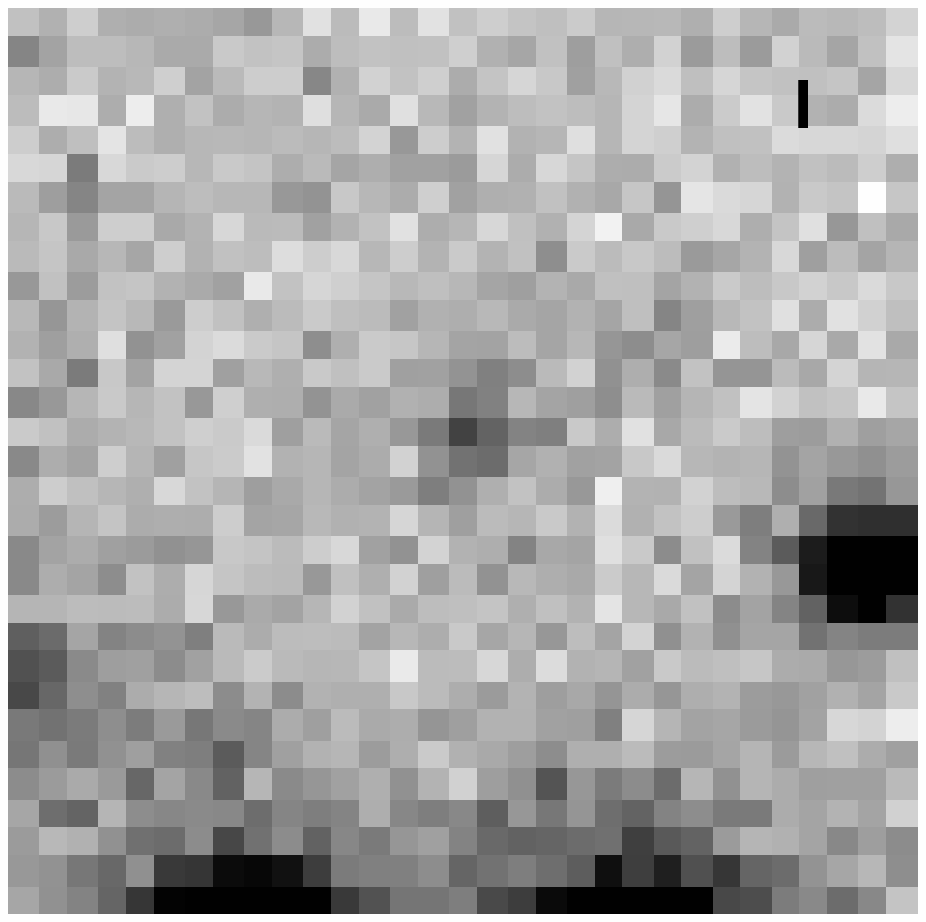}}
\put ( 38,38)   {\includegraphics[width=36mm,bb=175 275 435 535,clip]{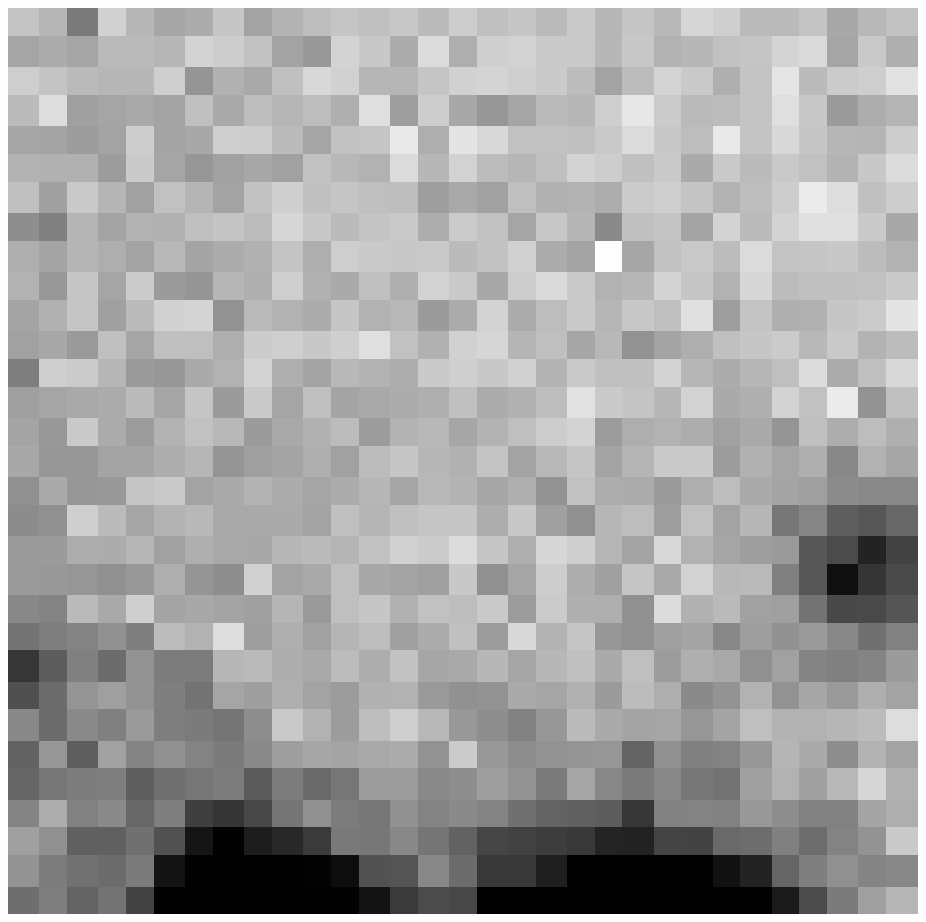}}
\put ( 38, 0)   {\includegraphics[width=36mm,bb=175 275 435 535,clip]{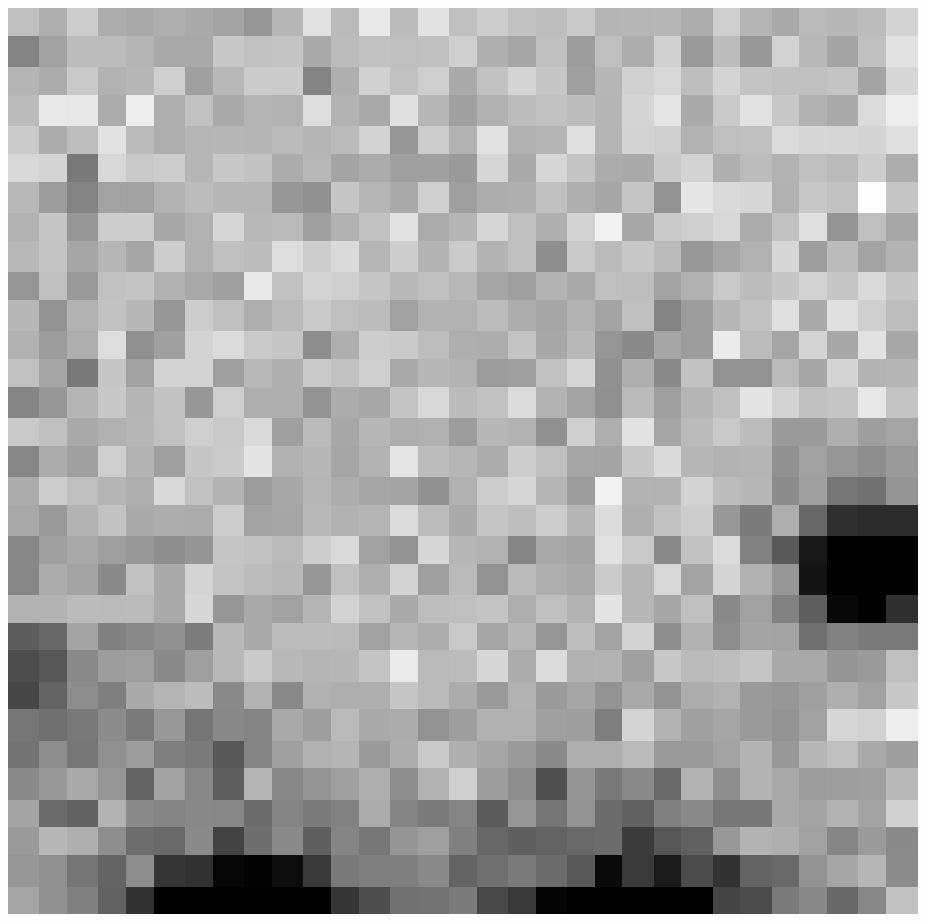}}
\end{picture}
\caption{$6\arcsec\times6\arcsec$ vicinity of the pulsar, as in small panels of 
 Fig.~\ref{f:ima},
obtained with the VLT/FORS1 in R and I bands (\cite{wagner}).
{\bfseries\itshape Left panels}
show original images. {\bfseries\itshape Right panels} show images with 
the modeled pulsar profile subtracted.}
\label{f:fors}
\end{figure}

\begin{figure}[t]
\setlength{\unitlength}{1mm}
\begin{picture}(80,40)(0,0)
\put (  0, 0)   {\includegraphics[width=36mm,bb=237 329 375 467,clip]{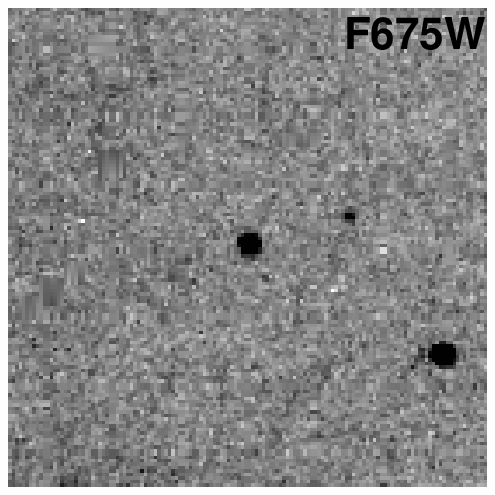}}
\put ( 38, 0)   {\includegraphics[width=36mm,bb=237 329 375 467,clip]{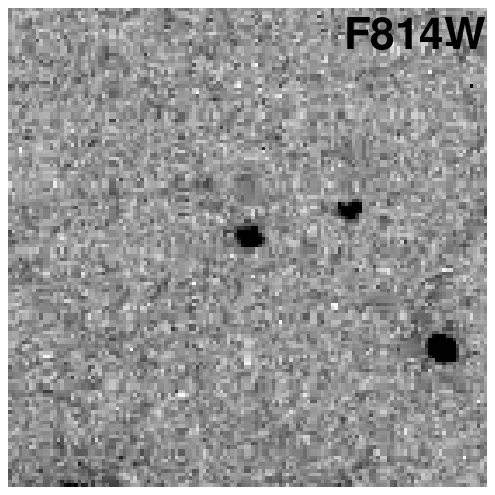}}
\end{picture}
\caption{$6\arcsec\times6\arcsec$ vicinity of the pulsar, as in small panels of 
 Figs.~\ref{f:ima}, and in Fig.~\ref{f:fors},
obtained with the \hst/WFPC2 in the F675W and F814W bands (\cite{mign}).}
\label{f:wfpc}
\end{figure}

 To search for the detected extended structures in the images 
 in adjacent bands we examined   
also the archival RI  band images of  the pulsar field, obtained 
with the VLT/FORS\footnote{Based 
on ESO programme 63.P-0002.} 
 on April 12, 1999 (\cite{wagner}),  
 and in the F675W (overlaps with R) and F814W (overlaps with I) bands,
 obtained with the \hst/WFPC2\footnote{Based on observations 
 made with the  NASA/ESA Hubble Space 
 Telescope, obtained from the data archive at the Space Telescope 
 Institute. STScI is operated by the association of Universities 
 for Research in Astronomy, Inc.\ under the NASA contract  NAS 5-26555.}
on March 19, 2000 and  on March 15, 2000, respectively
(\cite{mign}). 
The reduced images   are shown in Figs.~\ref{f:fors} and \ref{f:wfpc}. 
The pulsar is reliably detected in all 
bands, while the extended structures and o1 are not seen in any of them.
The extended structure near the pulsar could not be seen even after the pulsar PSF was
subtracted in the RI bands (Fig.~\ref{f:fors}).
 The object o2 is  seen in both \hst\ bands 
 (integrated exposure time 2600~s), but it is only barely 
  visible in the VLT I band and not detected in the short R band 
  exposure (300~s). This object and the extended 
  structures are also not visible in the \hst/WFPC2/F555W image 
(\cite{Caraveo}).
  This means that the detected extended 
  structures, as well as o1,  are red objects. 
  To conclude whether these objects are 
  associated with the pulsar nebula or 
they are background objects,
  additional observations are required. A change in
  brightness of these objects would strongly 
  support their association with the highly variable 
  structure of the PWN, as has been observed 
  in X-rays (\cite{pks01}). 
Note that wisp structures 
  have been detected and studied close to the Crab pulsar
  in X-rays   and in the optical   
(\cite{hest02}), and in the IR 
(\cite{sf02}). In X-rays and in the optical, the observations show that the
wisps vary in flux on a time scale of about one day.

\subsection{Photometry} 

\label{s:phot}

The photometric zeropoints for the ISAAC observations, 
$J_{\rm s0}=24.81(4)$ and $H_0=24.56(7)$, 
were derived using images of the standard 
star FS13, observed on December 15, 2000.
The J band catalog magnitude of FS13 (\cite{haw}) was used for the \js\ band. 
Differences in zeropoints between different ObsIDs were estimated 
comparing magnitudes of 6 field stars.
The differences proved to be no larger than 1\% in the \js\ and 2\% 
in the H bands, and were accounted for only in the resulting 
photometric error budget. 
The average Paranal atmospheric
extinction of 0.06 mag airmass$^{-1}$ was used in both bands\footnote{See 
\href{http://www.eso.org/instruments/isaac/imaging_standards.html}
{http://www.eso.org/instruments/isaac/imaging\_stan dards.html}}.
The fluxes of point-like objects  were measured in apertures with  
diameters  closest to mean seeing values (see Table~\ref{t:obs}),  
i.e., 4 pix in \js\ and 3 pix in H. 
Aperture corrections for the magnitudes were determined using 3 field stars. 
These stars, as well as the aperture diameters, were chosen to minimize 
the uncertainties of aperture corrected magnitudes of the faint pulsar, 
in both filters simultaneously.

An additional check of the photometry was performed using the 
IR-survey 2MASS\footnote{http://irsa.ipac.caltech.edu/}. Since 
the region of the Vela pulsar is not yet released in the archive catalog, 
we performed photometry of 10 stars present in the  J and H bands  
of the 2MASS ``Quicklook Images'' and in our ISAAC images. 
Their comparison shows that magnitude discrepancies are
$(J_{\rm s})_{\rm ISAAC}-J_{\rm 2MASS}=0.15(12)$ and
$H_{\rm ISAAC}-H_{\rm 2MASS}=0.03(10)$.
The discrepancy in the J band can be partly attributed to the difference
in throughput of the J and \js\ filters.
Although photometry on the 2MASS ``Quicklook Images'' is not recommended
because of their poor quality, we regard the obtained coincidence at 
the 1$\sigma$ level as 
a confirmation of the accuracy of   
our photometric referencing of the ISAAC data.

The pulsar magnitudes, determined with aperture photometry, 
are $J_{\rm s}=22.61(9)$ and $H=21.90(13)$. 
In addition, the pulsar profile was fitted with the PSF constructed  
 using {\tt IRAF/DAOPHOT}. 
The PSF-fitted magnitudes are $J_{\rm s}=22.71(10)$ and $H=22.04(16)$.
They differ by 1$\sigma$ from the aperture photometry magnitudes.
This difference may be attributed to the contamination of the pulsar counts 
measured with the aperture photometry by 
the extended structure  near the pulsar. 
We therefore consider the 
PSF-fitted magnitudes more reliable.
Using these magnitudes and the flux zeropoints by \cite{vdB},  
the measured fluxes from the pulsar  are 
$F_{\rm J_s}={1.39(12)}$~$\mu$Jy and $F_{\rm H}={1.64(25)}$~$\mu$Jy.  
We performed aperture photometry of the nearby objects 
o1, o2, and o3,  
which are marked in Fig.~\ref{f:ima}.  
The results of the photometry are summarized in Table~\ref{t:obj}.

We also measured the surface brightness of the extended source 
in the 1\farcs5\ vicinity of the pulsar. It was measured
on  the images with the pulsar subtracted over the area
(2.4 arcsec$^2$, the same in both bands)
which covers the brightest parts of
the structure. The surface brightnesses are
22.98(5) mag arcsec$^{-2}$ or 1.08(5) $\mu$Jy arcsec$^{-2}$,
and 21.51(7) mag arcsec$^{-2}$ or 2.66(16) $\mu$Jy arcsec$^{-2}$,
in the \js\ and H bands, respectively.
The surface brightness of the second structure  projected  
at the PWN inner arc (see Fig.~\ref{f:arc}) in the \js\ band is
25.8(7) mag arcsec$^{-2}$   
or  0.08(4) $\mu$Jy arcsec$^{-2}$, and the respective upper limit
in the H band is 24.44 mag arcsec$^{-2}$ or 0.18 $\mu$Jy arcsec$^{-2}$.

\begin{table*}
\caption{Photometry of the Vela pulsar (PSR) and  the
nearby objects o1, o2, and o3, marked in Fig.~\ref{f:ima}. 
 The offsets of the objects
from the pulsar position are given in the second column.  
Each spectral band cell for the PSR and o1  
consists of two pairs of magnitude/flux values, divided by lines:  
upper pair are measured values, lower pair are dereddened values.
Dereddening was performed with  $E_{B-V}=0.055(5)$.  
Each pair consists of the magnitude (upper value) and the flux
in $\mu$Jy (lower value).   
Only measured 
magnitudes/fluxes are presented for o2 and o3. 
All magnitudes are measured via aperture photometry,
except for the pulsar magnitudes in the \js\ and H bands, 
which are measured with PSF fitting  
(see Sect.~\ref{s:phot} for details). No magnitudes for \hst\ bands 
are calculated. Empty cells mean that the object is not measurable 
in this band.  
}
\label{t:obj}
\begin{tabular}{cccccccc}
\hline
\hline
Object & Offset & H & \js\ & I & \hst/F814W & R & \hst/F675W \\
\hline
PSR
& - &
\fourlines{22.04(18)}{1.64(25)}{22.00(18)}{1.69(26)} &  
\fourlines{22.71(10)}{1.39(12)}{22.66(10)}{1.45(13)} &  
\fourlines{22.95(13)}{1.57(18)}{22.86(13)}{1.71(19)} &  
\fourlines{-}{1.367(34)}{-}{1.497(38)} &                
\fourlines{23.42(11)}{1.29(12)}{23.29(11)}{1.46(14)} &  
\fourlines{-}{1.279(36)}{-}{1.456(45)} \\               
o1
& \twolines{0\farcs31 W}{1\farcs19 N} &
\fourlines{22.64(22)}{0.94(17)}{22.61(22)}{0.97(18)}    
& \fourlines{$\ge$24.10}{$\le$0.386}{$\ge$24.05}{$\le$0.403} 	
& - & \fourlines{}{$\le$0.131}{-}{$\le$0.144} & - &         
\fourlines{-}{$\le$0.130}{-}{$\le$0.148} \\                 
o2
& \twolines{1\farcs18 W}{0\farcs29 N} &
\twolines{20.81(11)}{5.07(47)}  &  
\twolines{21.94(7)}{2.83(18)} & 	
- & \twolines{-}{0.685(35)} & - &        
\twolines{-}{0.241(20)} \\               
o3
& \twolines{1\farcs14 W}{1\farcs67 S} &
\twolines{18.85(10)}{30.8(2.6)} &
\twolines{19.92(5)}{18.17(87)} &  
\twolines{21.53(4)}{5.84(21)} &    
\twolines{-}{5.361(59)} &                
\twolines{22.70(7)}{2.52(16)} &    
\twolines{-}{2.522(48)} \\                
\hline
\end{tabular}
\end{table*}

Since no photometric standards were observed
during the night of the VLT observations in the RI bands, 
the photometric equations 
were determined from
the Landolt standards\footnote{Fields of 
PG1323$-$085, PG1633$+$099, and PG1657$+$078 (\cite{Landolt}).}, 
observed on the night before:
\begin{eqnarray}
R-r = 26.583(09) +0.052(17) (r-i) \nonumber \\
I-i = 25.664(13) -0.068(23) (r-i) \nonumber
\end{eqnarray}
Here $R$ and $I$ are the Cousins magnitudes, $r$ and $i$ are 
the instrumental magnitudes. The ambient conditions 
monitor\footnote{Available at \href 
{http://archive.eso.org/asm/ambient-server}
{http://archive.eso.org/asm/ambient-server}}
shows that the average atmospheric extinction coefficient  
during the observations of 
the standards was in the range 0.140$-$0.155 mag/airmass, 
and at the time of the pulsar field observations it was
0.150 mag/airmass. Since the difference between them is 
negligible compared to the uncertainties of the equations presented above, 
we used these equations  for the photometric referencing of our observations 
without corrections for the extinction variations between the nights.    
The aperture corrections were done using a PSF constructed 
from 4 field stars. 
The measured pulsar magnitudes,    
$R = 23.46(11)$ and $I = 22.90(13)$, correspond to the fluxes
$F_{\rm R} = 1.24(12)$ $\mu$Jy and $F_{\rm I} = 1.65(19)$ $\mu$Jy,  
using the magnitude-flux conversion zero-points provided by  
\cite{Fkg}. 
The flux/magnitudes 
of the pulsar and o3 are shown in Table~\ref{t:obj}.

Pipeline-provided zeropoints and pivot wavelengths
were used for the flux calibration of the \hst\ observations 
($2.51\times10^{-18}$ erg cm$^{-2}$ s$^{-1}$ \AA$^{-1}$ / 7995 \AA\ and
$2.90\times10^{-18}$ erg cm$^{-2}$ s$^{-1}$ \AA$^{-1}$ / 6717 \AA\, 
in the F814W and F675W bands, respectively). 
Aperture photometry was performed for the pulsar and the objects 
o2 and o3.  
Aperture corrections were derived from a nearby relatively bright star. 
 The measured pulsar flux (see Table~\ref{t:obj}) 
in the F814W band is consistent with the published one (\cite{mign}), 
while the flux in the  F675W band is apparently   
$\simeq$25\% higher.  
Both fluxes are compatible with the  less accurate VLT fluxes 
in the RI bands described above.    
We measured also the fluxes of the nearby objects 
o2 and o3 and estimated  3$\sigma$ upper limits 
of the object o1 in the \hst\ bands.  

In  Table~\ref{t:obj} 
we also present dereddened magnitudes and fluxes for the pulsar and 
o1 using 
 $E_{B-V}=0.055(5)$ 
($A_{\rm V}\approx 0.18$, $R=3.1$). 
This corresponds to the column density  
$N_{\rm H}=3.3(3)\times 10^{20}$ cm$^{-2}$,
derived from the combined PL~$+$~NS-atmosphere spectral fit 
of the Vela pulsar X-ray data  obtained   
with the \chandra\ observatory (\cite{pzs01}).  
The extinction value is  consistent with the new distance 
to the Vela supernova remnant (SNR)
of  $250\pm30$~pc  (\cite{cha99}), 
and with the highest value over the Vela SNR 
$N_{\rm H}\simeq6 \times 10^{20}$~cm$^{-2}$ 
($A_{\rm V} \simeq 0.32$)  
found for its southern part
(\cite{luasc00}).

The colors of the brightest stellar object 
in the pulsar vicinity, o3, suggest that 
it could be a main sequence K5-K7 star at a distance 
of $\simeq 9$~kpc, assuming $A_{\rm V} \approx 2$. This is a much higher
extinction than we have adopted for the pulsar, but it is consistent with
the maximum possible Galactic extinction in the Vela direction, 
$A_{\rm V} \approx 4$ 
($N_{\rm H}\simeq 7.5 \times 10^{21}$~cm$^{-2}$; e.g., \cite{schleg98}). 
The colors of the fainter object o2 are roughly 
consistent with a cooler and even more distant main sequence star (of 
spectral type M0) at $\simeq 10-11$~kpc ($A_{\rm V} \approx 3$). 
The object o1 is too red to be consistent with any ordinary Galactic star.
It may be associated with the pulsar nebula. It may also be a background 
extragalactic object, as could also the objects o2 and o3. 
We discuss this further in Sect.~\ref{s:nearby}.

\section{Discussion}

\label{s:disc}

\subsection{Multiwavelength spectrum of the Vela pulsar}  

\begin{figure*}[t]
\includegraphics[width=150mm,clip]{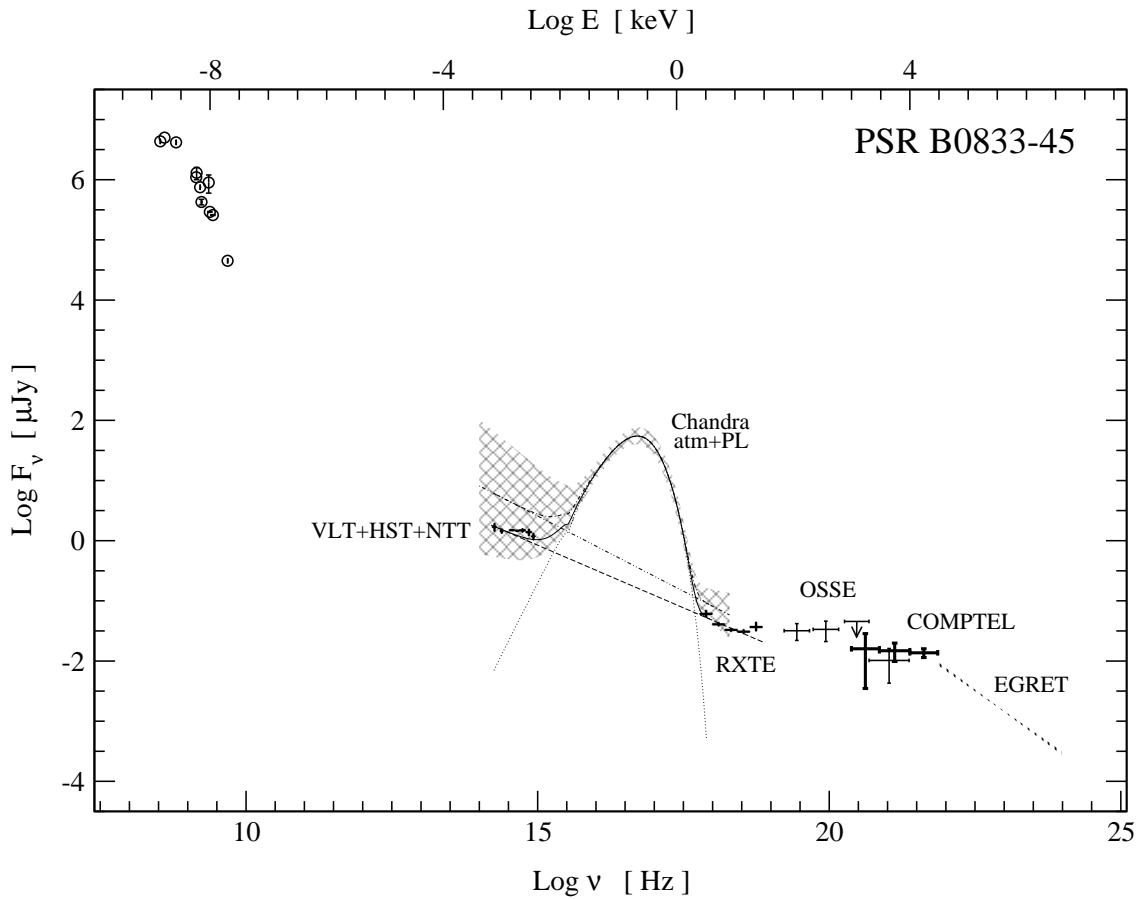}
\caption{Dereddened multiwavelength spectrum of the 
Vela pulsar obtained with different 
telescopes as 
marked in the plot.  Diamond-shaped fillings represent
$1\sigma$ confidence regions of the  
NS-atmosphere~$+$~PL fit (dot-dashed line)  
of the \chandra\ data; double-dot-dashed and dotted lines show 
the contributions of nonthermal PL and thermal atmosphere  
components, respectively.  The dashed line shows the PL fit 
of the near-IR and \rxte\ data and the solid line represents 
the sum of this PL component with the atmosphere component.    
The optical and \rxte\ ranges are shown enlarged in Fig.~\ref{f:spec}.
} 
\label{f:bspec}
\end{figure*}

 In Fig.~\ref{f:bspec}  we 
have 
combined our IR data with the available 
 phase-averaged  multiwavelength fluxes of the Vela pulsar including 
 the radio (EPN\footnote{European Pulsar Network (EPN) 
archive available at  
\href{http://www.mpifr-bonn.mpg.de/div/pulsar/data}
{http://www.mpifr-bonn.mpg.de/div/pulsar/data}.};   
S.~Johnston, 2002, private communications),
the optical 
(\cite{mign}), X-rays from the \chandra\  
(\cite{pzs01}),
 hard X-rays from the \rxte\ (\cite{hsg02}) and OSSE 
(\cite{str96}),
 and $\gamma$-rays from the COMPTEL 
(\cite{schon00})
and the EGRET  
(\cite{kan94}).
Here we present unabsorbed fluxes dereddened 
with  the color excess $E_{B-V}=0.055(5)$ applied to both the optical 
and X-ray regions. The VLT data in the RI bands 
are omitted since they are compatible with the more accurate 
\hst\ fluxes in the respective F675W and F814W bands (cf.\ Table~\ref{t:obj}).

While the pulsar flux generally decreases with increasing frequency,
one can resolve several nonthermal spectral components  
with different slopes in different spectral ranges.  
They are presumably of the pulsar magnetospheric origin.  
An excess in soft X-rays is 
attributed to the
thermal emission 
from the surface of the NS (\cite{og93}; \cite{pzs01}).  

\subsection{Phase averaged spectra of the pulsar in  
the optical and X-rays}   

As was noted by 
Pavlov et al.~(2001b), Mignani \& Caraveo (2001), 
Harding et al.~(2002), 
and seen from Fig.~\ref{f:bspec},  
the optical emission of the Vela pulsar is  
likely to be of nonthermal origin and the optical data  
are 
roughly 
compatible with   
the low energy extension of the X-ray PL 
spectral component dominating in the $2-10$~keV range.   
This may suggest a similar nature of the optical 
and the high energy X-ray emission. 
However, as seen from Fig.~\ref{f:bspec}, because of the limited  
statistics  of the available \chandra\ data in the $2-10$~keV range, 
the extension  
of the PL component inferred from the \chandra\ X-ray  
fit is much less certain  
than the near-IR and optical data. 

\begin{figure*}[t]
\includegraphics[width=170mm,bb=8 38 790 570,clip]{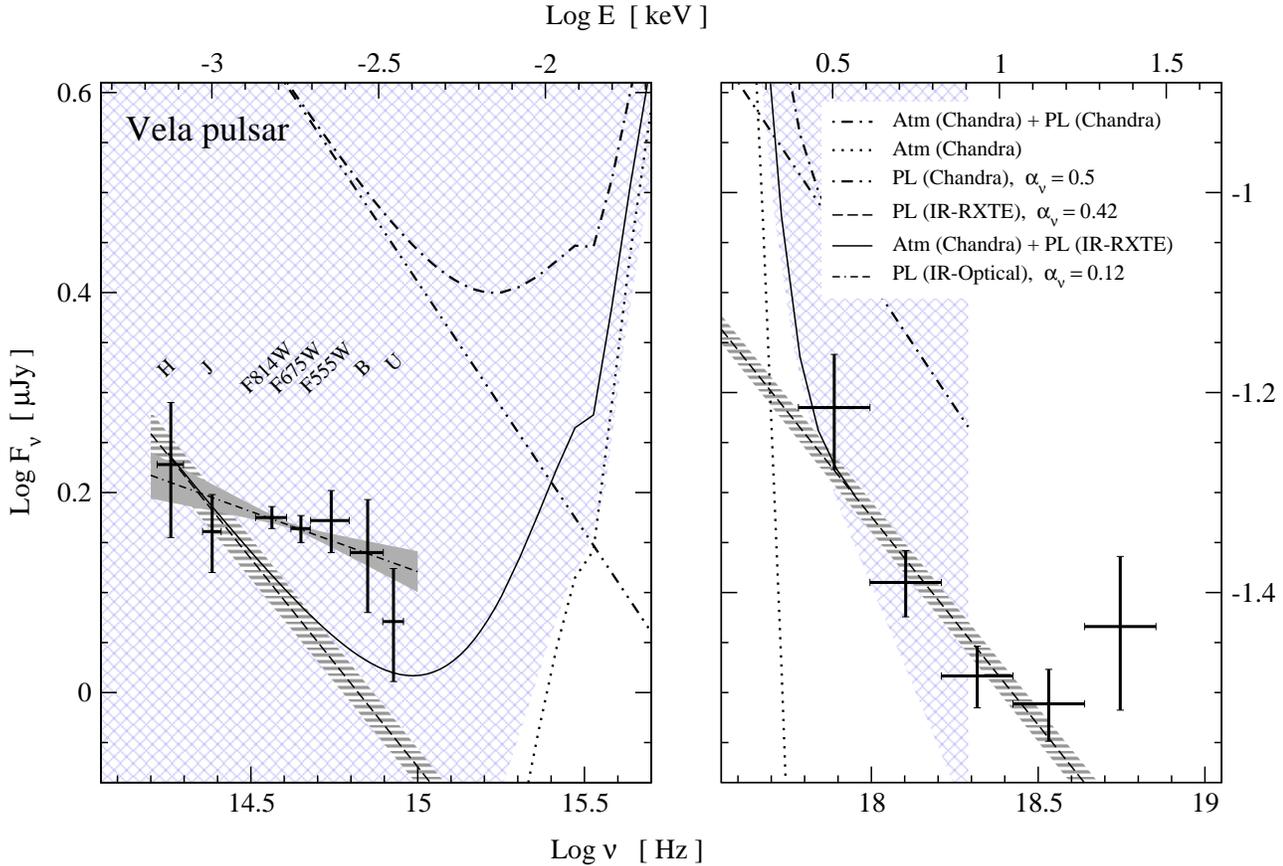}
\caption{Blow-up of the  
optical 
and X-ray   
parts of the Vela pulsar spectrum presented 
in Fig.~\ref{f:bspec}. The scale in both panels is the  same. 
{\bfseries\itshape Left panel:} IR-UV part of the pulsar
spectrum. IR and optical bands are marked.
Diamond-shaped, stripe-shaped and solid fillings represent
$1\sigma$ confidence regions of the NS-atmosphere~$+$~PL, 
IR-\rxte, and IR-Optical fits, respectively. The best fits 
and contributions of different spectral components are shown by different 
types of lines explained 
in the {\it right panel.\/} 
{\bfseries\itshape Right panel:} X-ray part of the spectrum.
Fluxes in the \rxte\ bands are marked by crosses. The flux in the hardest 
\rxte\ band likely belongs to a flat spectral component detected 
with the OSSE/COMPTEL  and was not used in the IR-\rxte\ fit. 
Line types are the same as  in the {\it left panel\/} and Fig.~\ref{f:bspec}.
}
\label{f:spec}
\end{figure*}  
  
At the same time, the \rxte\ data, being compatible 
with the \chandra\ results,  appear to be less uncertain. 
A difference in normalization of the PL components 
detected with \chandra\ and \rxte\ can be seen in Fig.~\ref{f:bspec}. 
It may be due to the fact that only the pulsed component 
is detected by \rxte\ (\cite{hsg02}).
But the \chandra\ observations, representing 
the total flux from the pulsar including an ``off-pulse'' component, 
show that the pulsed fraction 
in the \rxte\ range can be as high as 80\%  (\cite{san02})
and the  contribution of the off-pulsed component may  
not strongly affect the spectral shape.      
Excluding the 5th most energetic \rxte\ band, which is 
likely related to a flatter spectral component dominating  
in the OSSE and COMPTEL ranges (see Fig.~\ref{f:bspec}), 
 the \rxte\ spectrum can be  fitted with a PL with the spectral index 
$\alpha_{\nu}^{\rm RXTE} \simeq 0.41(9)$. Although this fit 
is  statistically inconsistent ($\chi^2=3.7$ per dof), 
it reflects the spectral slope in the \rxte\ range: 
 the best fit  line lies  within  a narrow  stripe-shaped region shown 
at the right panel of  Fig.~\ref{f:spec}.  
Its extension  to the optical range overlaps with the optical spectrum. 
A similar behavior was observed for the middle-aged pulsar PSR B0656$+$14  
(\cite{kopts01}). But the slope of the optical broad-band spectrum as a whole 
appears to be significantly  flatter, $\alpha_{\nu}^{\rm opt}=0.12(5)~(\chi^2=0.7$ per 
dof, see  the left panel of Fig.~\ref{f:spec}), and it is not possible to fit  
all optical and \rxte\  data with a single PL. We note also deviations from 
the single power law IR-Optical fit at about 1$\sigma$ level seen 
in the \js\ and U bands.

 Such behavior of the phase-averaged optical spectrum suggests that it 
can be a combination of several spectral components dominating at different 
phases of the pulsar light curve, as it is seen in the \rxte\ range (\cite{hsg02}).
This can only be proven by deep time-resolved photometry. 
To our knowledge, no such data have been obtained yet for the Vela pulsar. 
The most recent ``white-light'' time-resolved photometry reveals three peaks in the 
pulsar light curve in the optical range (\cite{gou98}).  
In contrast to that, up to 5 peaks were registered  
in the \rxte\ bands, and their PL spectra have significantly different 
indices and intensities. The second \rxte\ peak consists of two components, 
soft and hard, and coincides with the second optical peak (\cite{hsg02}). 
The second hard peak dominates the whole phase-averaged spectral flux, 
except for the 5th \rxte\ band, where the first peak with a positive 
spectral slope contributes significantly, providing 
a smooth connection to the spectral data in the OSSE range (cf.\ Fig.~\ref{f:bspec}). 
The first \rxte\ peak coincides with the first $\gamma$-ray peak.   

The measured fluxes in the \js\ and H bands may imply that 
the pulsar spectrum could be steeper in the IR than in the optical, 
as was also observed for 
the middle-aged pulsar PSR B0656$+$14  
(\cite{kopts01}).
Deeper observations of Vela in the H band are needed to state this 
possible similarity with greater confidence.
What is more obvious is that the spectral slope in 
the near-IR is   
compatible with the \rxte\ slope. Combining  
the first four \rxte\ bands with the \js\ and H bands gives a consistent 
PL fit with $\alpha_{\nu}^{\rm IR-RXTE} =0.417(6)$ 
($\chi^2=1.3$ per dof) shown by a dashed 
line in Figs.~\ref{f:bspec} and \ref{f:spec}. The optical bands show a flux 
excess over this fit. 

To better match the whole optical range  
we  combined this nonthermal component with the thermal 
NS atmosphere component describing the soft X-ray part 
of the \chandra\ data  
(\cite{pzs01},
combined HRC$+$ASIC fit).  
The combined model spectrum is shown by solid lines in Figs.~\ref{f:bspec} 
and \ref{f:spec}.
However, the atmosphere component decreases the residuals significantly 
only in the U and the softest \rxte\ bands, by contributions 
from the Rayleigh-Jeans and Wien tails of the thermal emission, 
respectively.        
The rest of the optical bands  still  show  
a significant excess over the combined atmosphere~$+$~PL model.  
Based on that, we can  speculate that the 2nd peak 
is responsible for the phase averaged emission  in both 
the \rxte\ and near-IR ranges, while the excess in BVRI is  
mainly produced by another phase and/or spectral component. 

Deeper \chandra\ observations of the Vela pulsar in the high energy  
tail of its X-ray emission 
are needed to perform more accurate 
phase averaged  spectral analysis of the optical and X-ray data.  
The high spatial resolution of \chandra\ 
should avoid the possible 
uncertainty 
of the \rxte\ fluxes which do not properly account for the off-pulsed 
component of the pulsar emission.
In this context, we can assume that the contribution of 
this component  just 
increases the \rxte\ fluxes 
by a factor of $\simeq 1.6$ (+0.2 in Log scale) 
in all bands to match the \chandra\ best 
PL fit line (double-dot-dashes in the {\it right panel\/} of Fig.~\ref{f:spec}).     
In this case, the low energy extension 
of the \rxte\ PL component would only overlap with the upper part of the 
B band error-bar, and would suggest a spectral break of the nonthermal 
component near the UB bands. On the other hand, if the current  
\chandra\ best PL fit is closer to reality, 
the break between 
the optical and X-ray slopes of the nonthermal component 
would be near 50 eV, i.e., in the EUV range. 
We consider these as alternative 
hypotheses to be tested by future observations.    
             
\begin{figure}[t]
\includegraphics[width=85mm,clip]{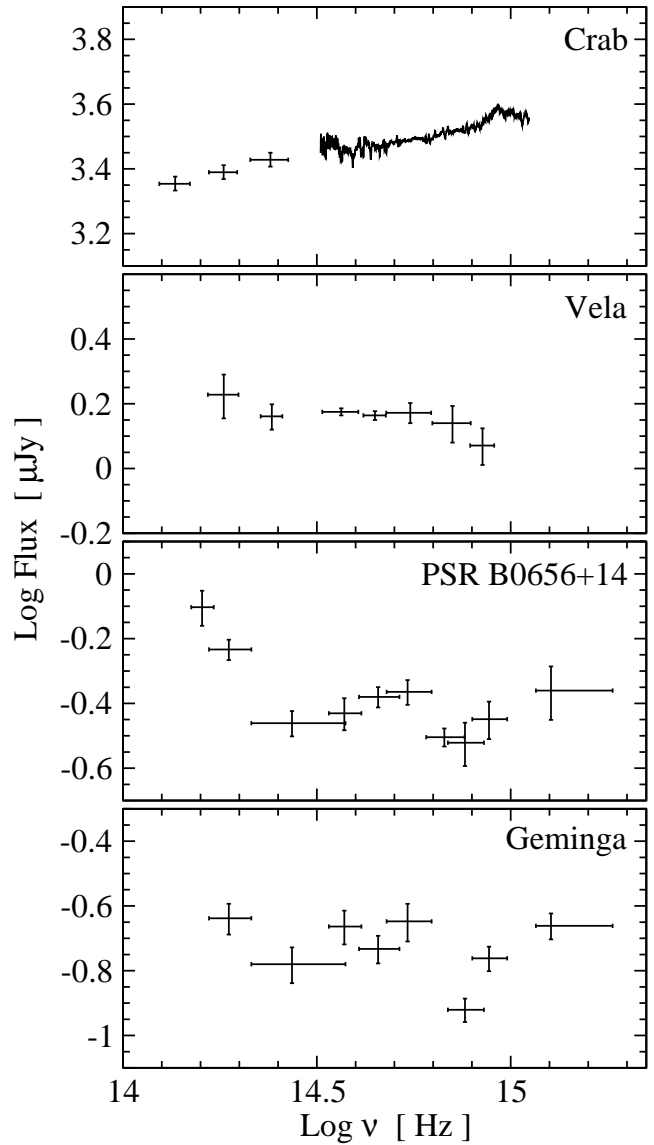}
\caption{Comparison of the optical spectra of four
pulsars. The youngest Crab 
(\cite{s03})
is at the {\it top\/},
the oldest Geminga   
(\cite{kom02})
is at the {\it bottom\/}.}
\label{f:spectra}
\end{figure}

\subsection{Optical spectrum of the Vela pulsar 
 and spectral evolution of the pulsar optical emission}

  The near-IR data extend significantly the broad-band optical 
  spectrum of the Vela pulsar towards longer wavelengths. This  
allows
a detailed comparison with the properties   
  of other pulsars observed in the near-IR.      
  In Fig.~\ref{f:spectra} we compare the optical-IR part of the spectrum  
  of the Vela pulsar with the available phase-averaged 
optical and near-IR spectra of other pulsars of different ages.  

Our photometry of the Vela pulsar in the \hst/F675W and VLT/R bands 
does not confirm a dip in  this range (\cite{mign}). 
Moreover, changing the extinction from $A_{\rm V}=0.4$ (\cite{man78})
to the most recent value $A_{\rm V}=0.18$     
   changes the average spectral index in the optical from $\alpha_\nu=-0.2(2)$
(\cite{mign}) to $\alpha_{\nu}^{\rm opt}=0.12(5)$. 
This is different from the most recently estimated positive 
slope of the spectrum of the younger ($\tau\approx 10^3$~yr) Crab pulsar, 
which shows
a monotonous flux increase from the IR to the FUV range  
(\cite{s00}; \cite{s03}).

Spectra of the middle-aged pulsars PSR B0656$+$14 and Geminga have dips 
 near around UB bands, and the spectrum of PSR B0656$+$14 has a strong 
increase in the near-IR.
We see signs of these features also in the Vela spectrum, although at a
low significance level.
 The spectrum of PSR B0656$+$14  increases significantly
towards the IR (\cite {kopts01}), contrary to the spectrum of the Crab pulsar.

\subsection{Extended structures}

\label{s:nearby}

Our IR observations  allowed us to detect  faint extended structures 
$\simeq$1\farcs5 SE and $\simeq3$\farcs1 NW of the pulsar.  
These are projected on the SE counter-jet and on the inner arc    
of the PWN detected in X-rays   
(\cite{hgh01}; \cite{pks01}).
  We also see a point-like object o1 $\simeq$1\farcs2 NW 
 of the pulsar, projected on the NW X-ray PWN jet.     
  All these objects are red and have no reliable counterparts 
 in the optical bands.  It is difficult to conclude 
whether  these features are associated with   
  the  PWN or they are background objects.

  Observations of the much more energetic Crab PWN show 
   that some of its structures are brighter in 
the IR than in the optical range.        For instance, the 
   knot structure, which is  only 0\farcs6 
   SE of the Crab pulsar, is more luminous  in the IR 
   and has a much steeper spectrum ($\alpha^{\rm knot}_{\nu} \simeq 0.8$)    
   than the Crab pulsar itself  (\cite{s03}).
This is consistent with the spatially averaged PL
   X-ray spectrum of the Crab PWN   (\cite{go02}).
   Some of the Crab wisps are also better resolved in the IR. 
 Based on that, and on the faintness of the Vela PWN as compared 
 with the Crab one, 
 it would not be too surprising if the Vela PWN  
 could be detected more easily in the near-IR than in the optical.

 In this context, the measured flux in the H band, and 
 our 3$\sigma$ detection  
 limits in the \js, \hst/F814W and \hst/F675W bands
 of the point-like  object o1 resembling the Crab knot, 
 suggest $\alpha^{\rm o1}_{\nu} \ga 1.8$.   
 Thus, its spectrum  is much steeper than that of the Crab knot  
 and also steeper than a spatially averaged X-ray spectrum  
 of the Vela PWN with   $\alpha^{\rm PWN}_{\nu} \simeq 0.5$  
(\cite{go02}). 
We are not aware of any reasonable
physical mechanism for   
 such a strong change of the spectral index   
 from X-rays to the optical range, and   
 we therefore believe that o1 is likely to be a distant highly absorbed 
 extragalactic object.  
    
\begin{figure}[t]
\includegraphics[width=88mm,clip]{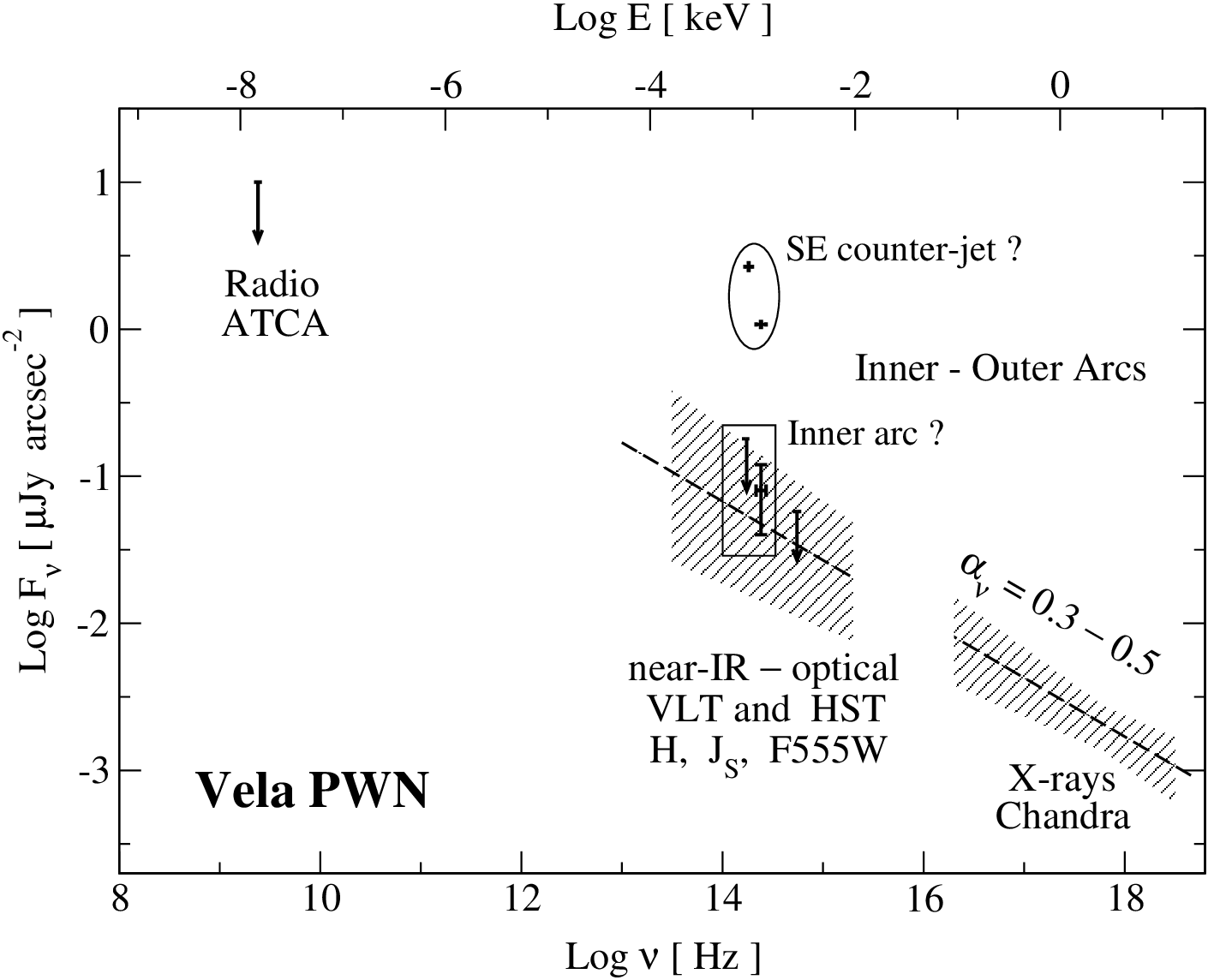}
\caption{Time and spatially averaged unabsorbed spectrum 
of the surface brightness of the Vela PWN 
inner and outer arc regions in X-rays fitted by a PL (\cite{kar02}) 
together with optical (\cite{mig02}) and radio (\cite{lew02})
upper limits, and the brightness of the 
suggested inner arc counterpart in the \js\ band and its upper limit in 
the H band marked by a box. 
Filled regions show uncertainties of the X-ray fit (dashed line) 
and its extension into the optical range.
The ellipse outlines the brightness of the near-IR extended structure 
projected at the SE X-ray counter-jet in $\simeq$1\farcs5  vicinity 
of the Vela-pulsar.  
}
\label{f:pwn}
\end{figure}

On the contrary,
as seen from Fig.~\ref{f:pwn}, 
the surface brightness in \js\ and its upper limit in H 
 of the IR inner arc shown in Fig.~\ref{f:arc} 
  are compatible with the low energy extension of the time 
  and spatially averaged X-ray  spectrum   
   of the Vela-PWN inner and outer   arcs. 
 The X-ray arc spectrum can be fitted with a PL 
 with $\alpha_\nu \simeq (0.3-0.5)$ and 
 unabsorbed  energy integrated brightness
  $\simeq (3-6) \times 10^{-14}$ ergs s$^{-1}$ cm$^{-2}$ arcsec$^{-2}$ 
  in the $0.1-10$~keV range  
(\cite{kar02}; \cite{mig02}).
  We consider this,  and the  
  positional coincidence of the inner IR and X-ray arcs as arguments 
in favor of the first detection of a counterpart of the Vela 
  PWN in the near-IR range. 
The IR brightness of the structure 
  is also consistent with the deepest optical upper limit of 0.057~$\mu$Jy 
  obtained recently in the \hst/F555W band (\cite{mig02}).
  If it is a real counterpart of the inner arc with the  spectral 
  slope described above,  only slightly deeper observations, presumably  
  at longer wavelengths, would allow a detection of this PWN structure.    
      
  The extended IR source closest to the pulsar, and apparently
  projected on the SE X-ray counter-jet, is an order of magnitude 
  brighter than expected from an extrapolation of the X-ray spectrum 
  into the near-IR/optical range (Fig.~\ref{f:pwn}). Its IR spectrum is 
  also much steeper, $\alpha_\nu \simeq 2.5$, 
  than the spectrum of the PWN further away from the pulsar.  
  As in the case of o1, this  
  does not argue in favor of it being associated
  with the PWN.
  It could, however, be that the inner jet structure 
  is brighter and has a steeper spectrum, because of possible 
  instabilities of the relativistic particle flow from the pulsar 
  as well as higher radiative losses at shorter distances from the pulsar.
  Although the upper limit on the optical flux from the IR source agrees
  with an extrapolation of the flux in the \js\ and H bands 
(see Fig.~\ref{f:pwn}), we
  emphasize that the optical and near-IR images are from different epochs.
  As mentioned in Sect.~\ref{s:astrmorph}, the emission of the regions
close to the Crab pulsar varies
  on a short time scale, and future comparison between optical and near-IR
  emission in the vicinity of the Vela pulsar would benefit from 
  simultaneous observations in these wavelength ranges.
        
 In X-rays the Vela PWN shows high variability    
 of its jet and arc-like structures   
 in  position, intensity, and hardness ratio  
(\cite{pks01}).
 Thus, further deep observations
of the pulsar field in the near-IR  and 
 in the optical would be useful to search for the variability and 
to prove or reject the association 
 of the detected extended structures  
 SE and NW of the pulsar and o1 with the PWN.  
 Observations of the Vela pulsar in the KL bands would be valuable to 
 investigate the possible increase of its flux towards the IR range.  
 Finally, time resolved photometry and  
spectral information
on the emission of different optical peaks of the pulsar pulse profile 
  would be crucial to understand  to which extent 
 nonthermal  optical radiation of the Vela pulsar 
is  of the  same origin as the nonthermal 
 spectral component seen in the high energy tail 
 of its X-ray spectrum, or whether it is generated by different 
 radiation mechanisms.

\section{Summary}

\label{s:sum}

Here we provide a summary of our most important results. 

\begin{list}{}{}
\item[1.] We have, for the first time, detected the Vela pulsar in the near-IR 
   in the \js\ and H bands. 

\item [2.] Our IR fluxes combined with the available broad-band optical data 
   confirm the nonthermal origin of the pulsar emission 
   in IR-optical range.    
   The combined  phase-averaged unabsorbed IR-optical 
   spectrum is fitted with a single PL with a negative slope.  
   This is in contrast to a positive slope of the unabsorbed 
   spectrum of the younger Crab pulsar.

\item [3.]  The  IR-optical spectrum and the phase-averaged
     PL spectral component  detected in the high energy tail        
     of the pulsar X-ray emission cannot be fitted with a single PL. 
     This suggests either a spectral break in the NUV-EUV range, 
	or the presence 
     of an additional spectral component with a flatter spectrum 
     dominating in the optical range. In the latter case,  
     the IR and the X-ray spectrum can be fitted  
     with a single PL suggesting the same  origin 
     of the nonthermal pulsar emission in the second 
     pulse of the pulsar pulse profile in both the X-ray and IR ranges.

\item [4.]   We detected two faint objects in the 
      1\farcs5  vicinity of the pulsar. They are projected on
     the SE counter-jet and the NW jet  of the Vela PWN detected in X-rays. 
      Both of them are extremely  red and have no counterparts 
      in the optical range. Their IR fluxes are 
    apparently inconsistent    
      with the expected IR brightness of 
      the PWN obtained by extrapolation of its X-ray spectrum into 
      the IR range. Finding variability of the objects would 
    strongly support their association with the highly variable PWN.

\item [5.]   A thin extended structure aligned with the inner arc 
      of the X-ray PWN is marginally seen in the \js\ band.  
      Its brightness is consistent with the X-ray PWN spectrum.  
      However, its reality and association with the PWN structure 
      need to be confirmed  at higher significance level 
      by deeper  observations.    
          
\end{list}

\acknowledgements{We are grateful to Soroush Nasoudi-Shoar 
for initial help with data reductions, to Simon Johnston for providing 
us with unpublished data on the radio spectrum of the Vela pulsar, 
to Alice Harding and Mark Strickman for tabulated results 
of the \rxte\ observations of the Vela pulsar, to Stefan Wagner 
for providing us with unpublished VLT data in the I band,    
to George Pavlov and Roberto Mignani 
for discussions and for access to the paper 
on the search for the optical counterpart of the  Vela PWN  prior to
publication, and to the referee Stephen Eikenberry for 
comments which allowed us to clarify better several points in text.
ABK and YAS are grateful
to Stockholm Observatory and the Royal 
Swedish Academy of Sciences, and ABK to the University of Washington,  
for hospitality. This work has been partially supported by the RFBR  
grants 02-02-17668, 03-02-17423, and 03-07-90200, 
the Royal Swedish Academy of Sciences and
the Swedish Research Council. PL is a Research Fellow at the Royal 
Swedish Academy of Sciences supported by a grant from the Wallenberg
Foundation.}


\begin{thebibliography}{}
 
\bibitem[Caraveo et al.\ 2001]{Caraveo}
Caraveo, P.\ A., De Luca, A., Mignani, R.\ P., et al. 
2001, ApJ, 561, 930

\bibitem[Cha et al.\ 1999]{cha99} 
Cha, A.\ N., Senbach, K.\ R., \& Danks, A.\ C. 
1999, ApJ, 499, L45

\bibitem [Cheng et al.\ 1986]{chr86} 
Cheng, K.\ S., Ho, C., \& Ruderman, M.\ A. 
1986, ApJ, 300, 500 
 
\bibitem [Daugherty \& Harding 1996]{dh96}  
Daugherty, J.\ K.\ \& Harding, A.\ K. 
1996, ApJ, 458, 278 

\bibitem[Fomalont et al.\ 1992]{Fom}
Fomalont, E.\ B., Goss, W.\ M., Lyne, A.\ G., et al. 
1992, MNRAS, 258, 497

\bibitem[Fukugita et al.\ (1995)]{Fkg} 
Fukugita, M., Shimasaku, K., \& Ichikawa, T. 
1995, PASP, 107, 945

\bibitem[Gotthelf \& Olbert 2002]{go02} 
Gotthelf, E.\ V.\ \& Olbert, C.\ M.
2002, in Proceedings of the 270. WE-Heraeus Seminar on Neutron Stars, Pulsars 
and Supernova Remnants, Physikzentrum Bad Honnef, Germany, Jan. 2002, 
eds.\ W.\ Becker, H.\ Lesch, \& J.\ Tr\"{u}mper, MPE Report 278, 159 
(astro-ph/02085169)

\bibitem[Gouiffes 1998]{gou98} 
Gouiffes, C. 
1998, in Neutron Stars and Pulsars, 
eds.\ N.\ Shibazaki, N.\ Kawai, S.\ Shibata \& T.\ Kifune 
(Univ. Acad. Press: Tokyo), 363 
 
\bibitem[Harding et al.\ 2002]{hsg02} 
Harding, A., Strickman, M.\ S., Gwinn, C., et al. 
2002 ApJ, 576, 376

\bibitem[Hawarden et al.\ 2000]{haw}
Hawarden, T.\ G., Leggett, S.\ K., Letawsky, M.\ B., et al.
2000, MNRAS, 325, 563

\bibitem[Helfand et al.\ 2001]{hgh01} 
Helfand, D.\ J., Gotthelf, E.\ V., \& Halpern, J.\ P. 
2001, ApJ, 556, 380 

\bibitem[Hester et al.\ 2002]{hest02} 
Hester, J.\ J.,  Mori, K.,  Burrows, D., et al.
2002, ApJ, 557, L49  
  

\bibitem[Kanbach et al.\ 1994]{kan94} 
Kanbach, G., Arzoumanian, Z., Bertsch, D., et al. 
1994, A\&A, 289, 855 

\bibitem[Kargaltsev et al.\ 2002]{kar02} 
Kargaltsev, O., Pavlov, G.\ G., Sanwal, D. et al. 
2002,  in Neutron Stars in Supernova Remnants, ASP Conf. Ser., 271, 
eds.\ P.\ O.\ Slane \& B.\ M.\ Gaensler (ASP: San Francisco), 181 
  
\bibitem[Komarova et al.\ 2002]{kom02} 
Komarova, V.\ N., Shibanov, Yu.\ A., Zharikov, S.\ V., et al. 
2002, in Proc. of the Workshop Pulsars, AXPs and 
SGRs observed with BeppoSAX and other observatories, in press    

\bibitem[Koptsevich et al.\ 2001]{kopts01} 
Koptsevich, A.\ B., Pavlov, G.\ G., Zharikov S.\ V., et al. 
2001, A\&A, 370, 304 

\bibitem[Labb\`{e}  et al. 2003]{labbe} 
 Labb\`{e}, I.,  Franx, M., Rudnick, G., et al. 2003, AJ, 125, 1107  

\bibitem[Landolt 1992]{Landolt} 
Landolt, A. 
1992, AJ, 104, 340

\bibitem[Lasker 1976]{las76} 
Lasker, B.\ M.
1976, ApJ, 203, 193 

\bibitem[Legge 2000]{leg00} 
Legge, D.
2000, in Pulsar Astronomy -- 2000 and Beyond, 
eds.\ M.\ Kramer, N.\ Wex, \& N.\ Welebinski, 
ASP Conference Series, 202, 141

\bibitem[Lewis et al.\ 2002]{lew02} 
Lewis, D., Dodson, R., McConnell, D., et al.
2002, in Neutron Stars and Supernova Remnants, ASP Conf. Ser., 271, 
eds.\ P.\ O.\ Slane \& B.\ M.\ Gaensler (ASP: San Francisco), 191 
  
\bibitem[Lu \& Aschenbach 2000]{luasc00} 
Lu, F.\ J.\ \& Aschenbach, B. 
2000, A\&A, 362, 1083

\bibitem[Manchester et al.\ 1978]{man78} 
Manchester, R.\ N., Lyne, A.\ G., Goss, W.\ M., et al.
1978, MNRAS, 184, 159 


\bibitem[Mignani \& Caraveo 2001]{mign} 
Mignani, R.\ P.\ \& Caraveo, P.\ A. 
2001, A\&A, 376, 213

\bibitem[Mignani et al.\ 2000]{Mignani2} 
Mignani, R.\ P., Caraveo, P.\ A., \& Bignami, G.\ F. 
2000, ESO Messenger, 99, 22

\bibitem[Mignani et al.\ 2003]{mig02} 
Mignani, R., De Luca, A., Kargaltsev, O., et al.
2003, A\&A, submitted 


\bibitem[\"{O}gelman et al.\ 1993]{og93} 
\"{O}gelman, H., Finley, J.\ R., \& Zimmerman, H.\ U. 
1993, Nature, 361, 136

\bibitem[Pavlov et al.\ 2001a]{pks01} 
Pavlov, G.\ G., Karagaltsev, O.\ Y., Sanwal, D., et al.
2001a, ApJL, 554, L189 



\bibitem[Pavlov et al.\ 2001b]{pzs01} 
Pavlov, G.\ G, Zavlin, V.\ E., Sanwal, D., et al. 
2001b, ApJL, 552, L129 

\bibitem[Romani 1996]{rom96} 
Romani, R.\ W. 
1996, ApJ, 470, 469 

\bibitem[Sanwal et al.\ 2001]{san02} 
Sanwal, D., Pavlov, G.\ G., Karagaltsev, O.\ Y., et al. 
2001. in Neutron Stars in SNRs, ASP Conference Series, 
eds.\ P.\ O. Slane \& B.\ M.\ Gaensler (astro-ph/0112164) 


\bibitem [Schlegel et al.\ 1998]{schleg98} 
Schlegel, D.\ J., Finkbeiner, D.\ P., Davis, M. 
1998, ApJ, 500, 525  


\bibitem[Sch\"{o}nfelder et al.\ 2000]{schon00} 
Sch\"{o}nfelder, V., Bennett, K., Blom, J.\ J., et al. 
2000, A\&AS, 143, 145 

 \bibitem[Simons  \&  Tokunaga \ 2002]{simons} 
 Simons, D. A.  \&  Tokunaga, A. 2002,
PASP, 114, 169  

\bibitem[Sollerman \ 2003]{s03} 
Sollerman, J. 2003, A\&A, submitted

\bibitem[Sollerman \& Flyckt 2002]{sf02} 
Sollerman, J.\ \& Flyckt, V. 
2002, ESO Messenger, 107, 32 
 
\bibitem[Sollerman et al.\ 2000]{s00} 
Sollerman, J., Lundqvist, P. Lindler, D., et al. 
2000, ApJ, 537, 861

\bibitem[Strickman et al.\ 1996]{str96} 
Strickman, M.\ S., Grove, J.\ E., Johnson, W.\ N., et al. 
1996, ApJ, 460, 735

\bibitem[Taylor et al.\ 1993]{Taylor}
Taylor, J.\ H., Manchester, R.\ N., \& Lyne, A.\ G. 
1993, ApJS, 88, 529

\bibitem[van der Bliek et al.\ (1996)]{vdB}
van der Bliek, N.\ S., Manfroid, J., \& Bouchet, P. 
1996, A\&ASS, 119, 547

\bibitem[Wagner \& Seifert 2000]{wagner} 
Wagner, S.\ J.\ \& Seifert, W. 
2000, in Pulsar Astronomy -- 2000 and beyond, ASP Conference series,
eds.\ M.\ Kramerr, N.\ Wex, \& N.\ Welebinski, 
 ASP Conference Series, 202,  315  


\bibitem[Wallace et al.\ 1977]{wal77}                             
Wallace, P.\ T., Peterson, B.\ A., Murdin, P.\ G., et al. 
1977, Nature, 266, 692

\end{thebibliography}
\end{document}